\documentclass[12pt]{elsart}
\usepackage{feynmp,epsfig,graphics,colordvi}
\def\be{\begin{eqnarray}}
\def\ee{\end{eqnarray}}
\def\bc{\begin{center}}
\def\ec{\end{center}}

\newcommand{\tr}{{\rm tr} \,}

\newcommand{\pslash}{\FMslash p}

\newcommand{\wslash}{\FMslash w}

\begin{document}
\begin{frontmatter}
\title{Open-charm systems in cold nuclear matter}
\author[GSI]{M.F.M. Lutz}
\author[PTE]{and C.L.\ Korpa}
\address[GSI]{Gesellschaft f\"ur Schwerionenforschung (GSI),\\
Planck Str. 1, 64291 Darmstadt, Germany}
\address[PTE]{Department of Theoretical Physics, University of
Pecs, \\Ifjusag u.\ 6, 7624 Pecs, Hungary}
\begin{abstract}
We study the spectral distributions of charmed meson with $J^P=0^-$ quantum numbers
in cold nuclear matter applying a self-consistent and covariant many-body approach
established previously for the nuclear dynamics of kaons. At leading orders the computation
requires as input the free-space two-body scattering amplitudes only. Our results are based
on the s-wave meson-nucleon amplitudes obtained recently in terms of a coupled-channel approach.
The amplitudes are characterized by the presence of many resonances in part so far not observed.
This gives rise to an intriguing dynamics of charmed mesons in nuclear matter. At nuclear saturation
density we predict a pronounced two-mode structure of the $D^+$ mesons with a main branch
pushed up by about 32 MeV. The lower branch reflects the coupling to two resonance-hole states
that are almost degenerate. For the $D^-$ we obtain a single mode pushed up by about 18 MeV relative
to the vacuum mode. Most spectacular are the results for the $D^+_s$ meson.
The presence of an exotic resonance-hole state gives rise to a rather broad and strongly momentum
dependent spectral distribution.
\end{abstract}
\end{frontmatter}

\section{Introduction}

It has been suggested to explore the possible mass modifications of D mesons with the
FAIR project at GSI \cite{FAIR}. It is planned to study such
effects by detecting D mesons  in heavy-ion reactions but also
in antiproton-nucleus reactions. The two approaches are complementary, with the second
focusing on the properties of D mesons in cold nuclear matter. Thus it is of importance
to develop a thorough theoretical understanding of how the D mesons may change their
properties in nuclear matter.

At present there are only few  works published on the properties of non-strange D mesons
\cite{Tsushima:1999,Sibirtsev:1999,Hayashigaki:2000,Tolos:Schaffner:Mishra:2004}.
To the best of our knowledge, there is no attempt to predict the properties
of the strange $D^\pm_s$  mesons in nuclear matter so far. Except for the work by Tolos et al.
\cite{Tolos:Schaffner:Mishra:2004} mean field parameterizations were considered.
The model by Tsushima et al. \cite{Tsushima:1999} arrived at attractive
mass shifts for the $D^+$ and $D^-$ mesons of about 60 MeV at saturation density.
A different mean field ansatz by Sibirtsev at al. \cite{Sibirtsev:1999} that considers also
vector mean field contributions suggests an attractive mass shift of about 140 MeV for the
$D^+$ meson, but a small repulsive mass shift of about 20 MeV for the $D^-$ meson.
A QCD sum rule analysis of Hayashigaki \cite{Hayashigaki:2000} predicted
an attractive mass shift of about 50 MeV for the $D^+ $ meson. These
developments resemble to some extent the first attempts to predict the
properties of kaons and antikaons in nuclear matter (see e.g.
\cite{Kaplan:Nelson,njl-lutz:a,njl-lutz:b}),
which also assumed a mean field type behavior for the mass shifts in nuclear matter.
However, by now it is well established that for the antikaon such an ansatz is not valid due to
complicated many-body dynamics induced by the presence of resonance-hole states
(see e.g. \cite{ml-sp,Lutz:Korpa:2002}). Thus it may be crucial to study also the properties
of D mesons in a more microscopic manner.

An important step towards a more realistic approach to the $D^+$ meson in nuclear matter was taken
by Tolos et al. \cite{Tolos:Schaffner:Mishra:2004}, who did not insist on a mean field approach
exploring the possible influence of the $\Lambda_c(2594)$ resonance on the nuclear dynamics of the
$D^+$ meson. The key ingredient of a realistic description are the $D^+ N\to D^+ N$ scattering amplitudes,
that determine the $D^+$ meson self energy at least for dilute nuclear matter
\cite{dover,njl-lutz:b}. In turn  the results of \cite{Tolos:Schaffner:Mishra:2004} depend
decisively on their particular form of the coupled-channel amplitudes that took into account
the $ \pi \Sigma_c ,D N, \eta \Lambda_c$ channels for $I=0$ and
$  \pi \Lambda_c, \pi \Sigma_c, D N,  \eta \Sigma_c$ channels for $I=1$. The $\Lambda_c(2594)$
resonance which carries $J^P=\frac{1}{2}^-$ quantum numbers
was dynamically generated in \cite{Tolos:Schaffner:Mishra:2004} as suggested first
in \cite{Lutz:Kolomeitsev:2004}. The interaction strengths were derived relying on a
SU(3) symmetry for the u,d and c quarks. As a consequence of their amplitudes an attractive mass shift
of about 10 MeV together with a significant broadening of the $D^+$ is predicted at nuclear
saturation density was predicted. Recently a more complete computation \cite{Hofmann:Lutz:05}
that incorporated the additional channels $K \Xi_c, K \Xi_c', D_s \Lambda$ for $I=0$ and
$K \Xi_c, K \Xi_c', D_s \Sigma$ for $I=1$ arrived at scattering amplitudes that differ
significantly from those of \cite{Tolos:Schaffner:Mishra:2004}. The interaction was saturated
by a t-channel vector meson exchange. Comparing the different interactions
\cite{Tolos:Schaffner:Mishra:2004,Hofmann:Lutz:05} the work by Tolos et al.
severely overestimates the charm-exchange channels. In \cite{Hofmann:Lutz:05} those
channels are suppressed by a kinematical factor $m^2_\rho/m^2_D \sim 0.2 $.
As an important new result of \cite{Hofmann:Lutz:05}
the $I=0$ amplitude shows two resonance structures around 2.6 GeV. A narrow state that couples
dominantly to the $D N, D_s \Lambda$ channels and a broader state that is interpreted as a chiral
excitation of the open-charm sextet $\frac{1}{2}^+$ ground states \cite{Lutz:Kolomeitsev:2004}. The
narrow state should be identified with the observed state $\Lambda_c(2594)$. The broader one, which
couples strongly to the $\pi \Sigma_c$ channels, awaits experimental confirmation and couples very
weakly to the $D N$ channel. It is the analogue of the $\Lambda(1405)$, which is a chiral excitation
of the baryon octet $\frac{1}{2}^+$ ground states (see e.g. \cite{Wyld,Dalitz,Granada,Copenhagen}).
In contrast to \cite{Tolos:Schaffner:Mishra:2004} the isospin one
amplitude of \cite{Hofmann:Lutz:05} reflects a narrow resonance of mass 2.62 GeV which couples
dominantly to the $D N, D_s \Sigma$ channels. This additional resonance will affect the
properties of $D^+$ mesons in nuclear matter significantly.

The purpose of this letter is threefold. First we derive the
properties of the $D^+$ meson in nuclear matter based on the improved understanding of the
$D^+ N$ scattering amplitudes of \cite{Hofmann:Lutz:05}. Second, we present the mass shift
of the $D^-$ as predicted by the scattering amplitudes of \cite{Hofmann:Lutz:05}. Thirdly, we
present for the first time predictions on how the strange $D^\pm_s$ mesons change their
properties in a dense nuclear environment. Again the scattering amplitudes as obtained in
\cite{Hofmann:Lutz:05} are used. The s-wave $D^\pm_s N \to D^\pm_s N$
scattering amplitudes are characterized by exotic resonances at 2.89 GeV and 2.78 GeV.
The computations are based on the self consistent and covariant
many-body approach developed in \cite{ml-sp,Lutz:Korpa:2002}. It is important to perform such
computations in a self consistent manner since the feedback of an altered meson spectral function
on the resonance structure proved to be a decisive many-body affect \cite{ml-sp,Lutz:Korpa:2002,Korpa}.

\section{Self consistent and covariant nuclear dynamics for charmed mesons}

In this section we recall the self consistent and relativistic many-body framework
required for the evaluation of the pseudo-scalar meson propagation in nuclear matter
\cite{Lutz:Korpa:2002}. The key ingredient are the vacuum on-shell meson-nucleon scattering
amplitudes
\begin{eqnarray}
\langle D^{j}(\bar q)\,N(\bar p)|\,T\,| D^{i}(q)\,N(p) \rangle
&=&(2\pi)^4\,\delta^4(q+p-\bar q-\bar p )\,
\nonumber\\
&& \!\!\!\!\!\times \,\bar u(\bar p)\,
T^{ij}_{D N \rightarrow D N}(\bar q,\bar p ; q,p)\,u(p) \,,
\label{on-shell-scattering}
\end{eqnarray}
where  $\delta^4(..)$ guarantees energy-momentum conservation and $u(p)$ is the
nucleon isospin-doublet spinor. Here is $D =(D_+, D_0), ( \bar D_0, D_-), D^+_s $ or $D^-_s$
depending on which system we study. The scattering amplitudes are decomposed into their
isospin subsystems. The reactions  $D^\pm_s N \to D^\pm_s N$ carry $I=1/2$ inheriting
the isospin of the nucleons. The amplitudes involving the isospin doublets $(D_+, D_0)$ or
$( \bar D_0, D_-)$  are decomposed into isospin
zero and one components
\begin{eqnarray}
&&T^{i j}_{D N \to D N}(\bar q,\bar p \,; q,p)
= T^{(0)}_{D N\to D N}(\bar k,k;w)\,P^{ij}_{(I=0)}+
T^{(1)}_{D N\to D N}(\bar k,k;w)\,P^{ij}_{(I=1)}\;,
\nonumber\\
&& P^{ij}_{(I=0)}= \frac{1}{4}\,\Big( \delta^{ij}\,1
+ \big( \vec \tau\,\big)^{ij}\,\vec \tau \,\Big)\, , \qquad
P^{ij}_{(I=1)}= \frac{1}{4}\,\Big( 3\,\delta^{ij}\,1-
\big(\vec \tau \,\big)^{ij}\,\vec \tau \,\Big)\;,
\label{}
\end{eqnarray}
where $q, p, \bar q, \bar p$ are the initial and final meson and nucleon 4-momenta and
\begin{eqnarray}
w = p+q = \bar p+\bar q\,,
\quad k= \half\,(p-q)\,,\quad
\bar k =\half\,(\bar p-\bar q)\,.
\label{def-moment}
\end{eqnarray}

The in-medium meson-nucleon scattering amplitude is determined by the coupled-channel
Bethe-Salpeter equation:
\begin{eqnarray}
{\mathcal T}(\bar k ,k ;w ) &=& {\mathcal K}(\bar k ,k ;w )
+\int \frac{d^4l}{(2\pi)^4}\,{\mathcal K}(\bar k , l;w )\, {\mathcal G}(l;w)\,{\mathcal T}(l,k;w )\;,
\label{hatt}
\end{eqnarray}
where the  in-medium scattering amplitude ${\mathcal T}(\bar k,k;w,u)$ and the two-particle
propagator ${\mathcal G}(l;w,u)$ depend on the 4-velocity $u_\mu$
characterizing the nuclear matter frame. For nuclear matter moving with a velocity
$\vec v$ one has
\begin{eqnarray}
u_\mu =\left(\frac{1}{\sqrt{1-\vec v\,^2/c^2}},\frac{\vec v/c}{\sqrt{1-\vec v\,^2/c^2}}\right)
\;, \quad u^2 =1\,.
\label{}
\end{eqnarray}
We emphasize that (\ref{hatt}) is properly defined from a Feynman diagrammatic point of
view even in the case
where the in-medium scattering process is no longer well defined due to a broad meson
spectral function.

To be consistent with the dynamics derived in \cite{Hofmann:Lutz:05} all channels must be considered
in (\ref{hatt}) that were incorporated in \cite{Hofmann:Lutz:05}. In this work we exclusively
study the effect of an in-medium modified two-particle propagator, i.e. we identify the in-medium
interaction kernel ${\mathcal K}$ in (\ref{hatt}) with its vacuum limit to be taken from
\cite{Hofmann:Lutz:05}. We restrict ourselves
further to the in-medium change of the D mesons only, i.e. all two-particle propagators in
(\ref{hatt}) that do not involve a D meson are taken to be of the form used
in \cite{Hofmann:Lutz:05}. We write
\begin{eqnarray}
&& \!\!\!\!\Delta S (p,u) = 2\,\pi\,i\,\Theta \Big(p\cdot u \Big)\,
\delta(p^2-m_N^2)\,\big( \pslash +m_N \big)\,
\Theta \big(k_F^2+m_N^2-(u\cdot p)^2\big)\,,
\nonumber\\
&&\!\!\!\!{\mathcal S}(p,u) = S(p)+ \Delta S(p,u)\,, \quad
{\mathcal D}(q,u)=\frac{1}{q^2-m_D^2-\Pi(q,u)} \;,
\nonumber\\
&& \!\!\!\!{\mathcal G}_{DN}(l;w,u) = -i\,{\mathcal S}({\textstyle
{1\over 2}}\,w+l,u)\,{\mathcal D}({\textstyle {1\over 2}}\,w-l,u)  \;,
\label{hatg}
\end{eqnarray}
where the Fermi momentum $k_F$ parameterizes the nucleon density $\rho$ with
\begin{eqnarray}
\rho = -2\,\tr \,\gamma_0\,\int \frac{d^4p}{(2\pi)^4}\,i\,\Delta S_N(p,u)
= \frac{2\,k_F^3}{3\,\pi^2\,\sqrt{1-\vec v\,^2/c^2}}  \;.
\label{rho-u}
\end{eqnarray}
In the rest frame of the bulk with $u_\mu=(1,\vec 0\,)$ one recovers with (\ref{rho-u}) the
standard result $\rho = 2\,k_F^3/(3\,\pi^2)$. In this work we also refrain
from including nucleonic correlation effects. The meson self energy $\Pi(q,u)$ is evaluated self
consistently in terms of the in-medium scattering amplitudes
\begin{eqnarray}
&&\Pi(q,u) = 2\,\tr \int \frac{d^4p}{(2\pi)^4}\,i\,\Delta S(p,u)\,
\bar {\mathcal T}\big({\textstyle{1\over 2}}\,(p-q),
{\textstyle{1\over 2}}\,(p-q);p+q,u \big)\,,
\nonumber\\
&& \bar {\mathcal T}= \frac{1}{4}\,{\mathcal T}_{D N\to D N}^{(I=0)}+
\frac{3}{4}\,{\mathcal T}_{D N \to D N}^{(I=1)} \;, \qquad {\rm or} \quad
\bar {\mathcal T}= {\mathcal T}^{(I=\frac{1}{2})}_{D^\pm_s N \to D_s^\pm N} \,.
\label{k-self}
\end{eqnarray}
In order to solve the self consistent set of equations (\ref{hatt},\ref{hatg},\ref{k-self})
it is convenient to rewrite the coupled-channel system.
Given our assumptions
the coupled-channel problem reduces without any further restrictions to a single channel problem:
\begin{eqnarray}
&&{\mathcal T}_{D N \to D N}^{(I)}=  T_{D N \to D N}^{(I)}
+T_{D N \to D N}^{(I)}\cdot \Big({\mathcal G}_{D N}-G_{D N} \Big)
\cdot {\mathcal T}_{D N \to D N}^{(I)}\;,
\label{rewrite:b}
\end{eqnarray}
where we use a compact notation.
The self consistent set of equations (\ref{hatg},\ref{k-self},\ref{rewrite:b})
is now completely determined by the vacuum amplitudes $T_{D N \to D N}^{(I)}$.
Rather than specifying the coupled-channel interaction kernel of \cite{Hofmann:Lutz:05}
it is sufficient to recall the s-wave scattering matrix describing  the $D N \to DN$ process.
The one of \cite{Hofmann:Lutz:05} has the separable form
\begin{eqnarray}
&& T^{(I)}(w) = \frac{1}{2}\,\left( \frac{\wslash}{\sqrt{w^2}}+ 1 \right)M^{(I)}(\sqrt{s}\,)\,,
\nonumber\\
&&f^{(I)}(\sqrt{s}\,) = \frac{1}{8\,\pi\,\sqrt{s}}
\left( \frac{\sqrt{s}}{2}+\frac{m_N^2-m_D^2}{2\,\sqrt{s}} +
m_N\right) M^{(I)}(\sqrt{s}\,)\,
\nonumber\\
&& \qquad \qquad  \,=
\frac{1}{2\,i\,p_{D N}}\left(\eta^{(I)}(\sqrt{s}\,)\,
e^{2\,i\,\delta^{(I)} (\sqrt{s}\,)}-1 \right) \;,
\nonumber\\
&& \qquad \!\frac{p_{D N}^2}{s} = \frac{1}{4}\,\Big(1-
\frac{(m_N+m_D)^2}{s} \Big)\,\Big(
1-\frac{(m_N-m_D)^2}{s}\Big) \,,
\label{t-vacuum}
\end{eqnarray}
where we recall the parametrization of the amplitudes in terms of phase shift $\delta $ and inelasticity $\eta $.

Using (\ref{t-vacuum}) as input in (\ref{rewrite:b}) the in-medium amplitude takes the form
\begin{eqnarray}
&&{\mathcal T}^{(I)}(w,u) =
\frac{1}{2}\,\left( \frac{\wslash}{\sqrt{w^2}}+ 1 \right){\mathcal M}^{(I)}(w,u)\,,
\nonumber\\
&& {\mathcal M}^{(I)}(w,u) = \Big[ 1 -M^{(I)}(\sqrt{s}\,)\,\Delta J(w,u)\Big]^{-1}
M^{(I)}(\sqrt{s}\,)  \;,
\label{m:eq}
\end{eqnarray}
where we neglect contributions of higher partial-wave amplitudes. The p-wave and d-wave
scattering amplitudes are not determined in \cite{Hofmann:Lutz:05}.
The reduced loop function $\Delta J(w,u)$ acquires the generic form
\begin{eqnarray}
&&\Delta J(w,u) = \int \frac{d^4l}{(2 \pi)^4}\,g(l;w,u)\,
\left(m_N+ \frac{l \cdot w}{\sqrt{w^2}}\right)\,,
\nonumber\\ \nonumber\\
&& g(l;w,u) = 2\,\pi\,\Theta \Big(l\cdot u \Big)\,\delta(l^2-m_N^2)\,
\frac{\Theta \Big(k_F^2+m_N^2-(u \cdot l)^2 \Big)}{(w-l)^2-m_D^2-\Pi(w-l,u)}
\nonumber\\
&& \qquad \qquad -\frac{i}{l^2-m_N^2+i\,\epsilon} \,
\frac{1}{(w-l)^2-m_D^2-\Pi(w-l,u)}
\nonumber\\
&& \qquad \qquad +\frac{i}{l^2-m_N^2+i\,\epsilon} \,\frac{1}{(w-l)^2-m_D^2+i\,\epsilon}\;.
\label{j-exp}
\end{eqnarray}
We observe
that the loop function $\Delta J(w,u)$ is a scalar under Lorentz transformation and therefore
depends only on $w^2$ and $w \cdot u$. Thus it can be evaluated in any
convenient frame without loss of information. In practice we perform the loop integration
in the rest frame of nuclear matter with $u_\mu=(1,\vec 0)$. It is straightforward
to perform the energy and azimuthal angle integration in (\ref{j-exp}). The energy integration
of the last two terms in (\ref{j-exp}) is performed by closing the complex contour in the
lower complex half plane. One picks up two contributions: first the nucleon pole leading
to $l_0=\sqrt{m_N^2+\vec l^2}$ in (\ref{j-exp}) and second the meson pole at typically
$w_0-l_0 = -\sqrt{m_D^2+(\vec l-\vec w\,)^2}$. Here we neglect the meson pole contribution.
All together one is left with a two-dimensional integral which must
be evaluated numerically. In our simulation we restrict $|\vec l\,| < 800$ MeV.

The meson self energy follows
\begin{eqnarray}
&&\Pi(q,u) = -2\,\int_0^{k_F} \frac{d^3 p}{(2\pi)^3} \,
\left(\frac{m_N}{E_p}+ \frac{w_0^2-\vec w\,^2-q\cdot w}{E_p\,\sqrt{w^2}}\right)\,\bar {\mathcal M}(w,u) \,,
\nonumber\\
&&\bar {\mathcal M}(w,u) = \frac{1}{4}\,{\mathcal M}^{(I=0)}(w,u)
+\frac{3}{4}\,{\mathcal M}^{(I=1)}(w,u)\,,
\label{kaon-final}
\end{eqnarray}
where we used $u_\mu=(1, \vec 0\,), q_\mu=(\omega, \vec q\,)$ and
$w_\mu = (\omega+E_p, \vec q+\vec p\, )$ with $E_p=(m_N^2+\vec p\,^2)^{1/2}$.
With  (\ref{m:eq}), (\ref{j-exp}) and (\ref{kaon-final}) a
self consistent set of equations is defined. It is solved numerically by iteration. First
one determines the leading meson self energy $\Pi(\omega ,\vec q\,)$
by (\ref{kaon-final}) with ${\mathcal M}(w,u) =M(\sqrt{s})$. That leads via the loop
functions (\ref{j-exp}) and the in-medium Bethe-Salpeter equation (\ref{m:eq}) to medium
modified scattering amplitudes ${\mathcal M}(w_0,\vec w\,) $. The latter are used
to determine the meson self energy of the next iteration. This procedure typically
converges after 3 to 4 iterations. The manifest covariant form of the self
energy and scattering amplitudes are recovered with
$\Pi(q^2,\omega)= \Pi(q^2,q \cdot u)$
and ${\mathcal M}(w^2,w_0)={\mathcal M}(w^2,w \cdot u)$ in a straightforward
manner if considered as functions of $q^2,\omega $ and $w^2, w_0$ respectively.

\newpage

\section{Results}

\begin{figure}[b]
\begin{center}
\includegraphics[width=14.5cm,clip=true]{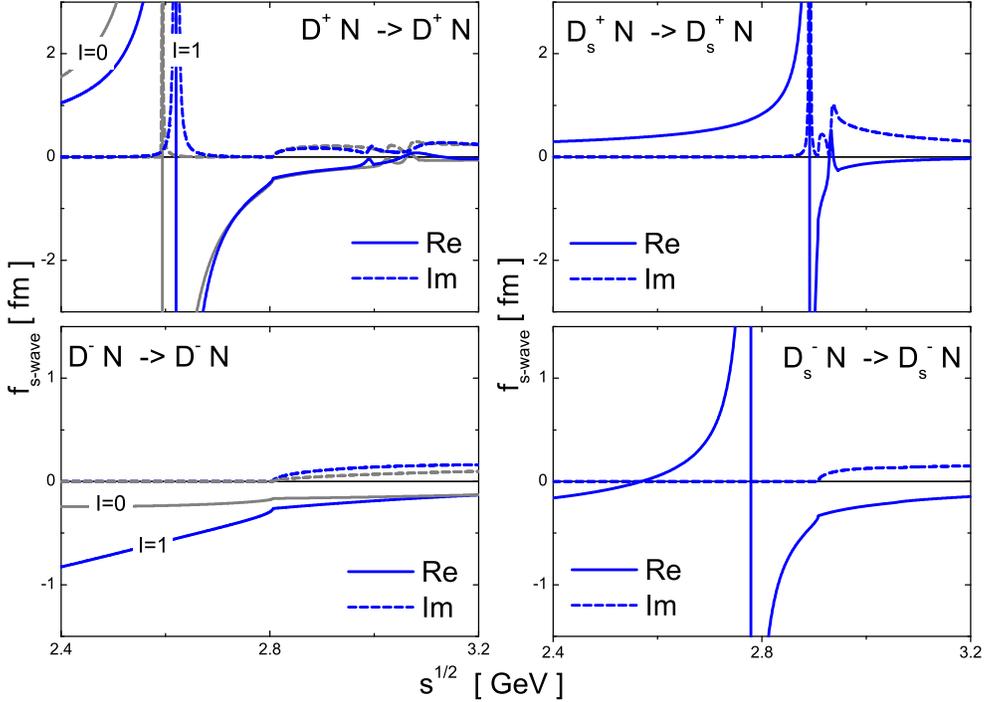}
\end{center}
\caption{S-wave scattering amplitudes of the D mesons off the nucleon. The amplitudes
are taken from \cite{Hofmann:Lutz:05}. The normalization of the amplitudes is specified in
(\ref{t-vacuum}).}
\label{fig:amplitudes}
\end{figure}

\begin{figure}[t]
\begin{center}
\includegraphics[width=14.5cm,clip=true]{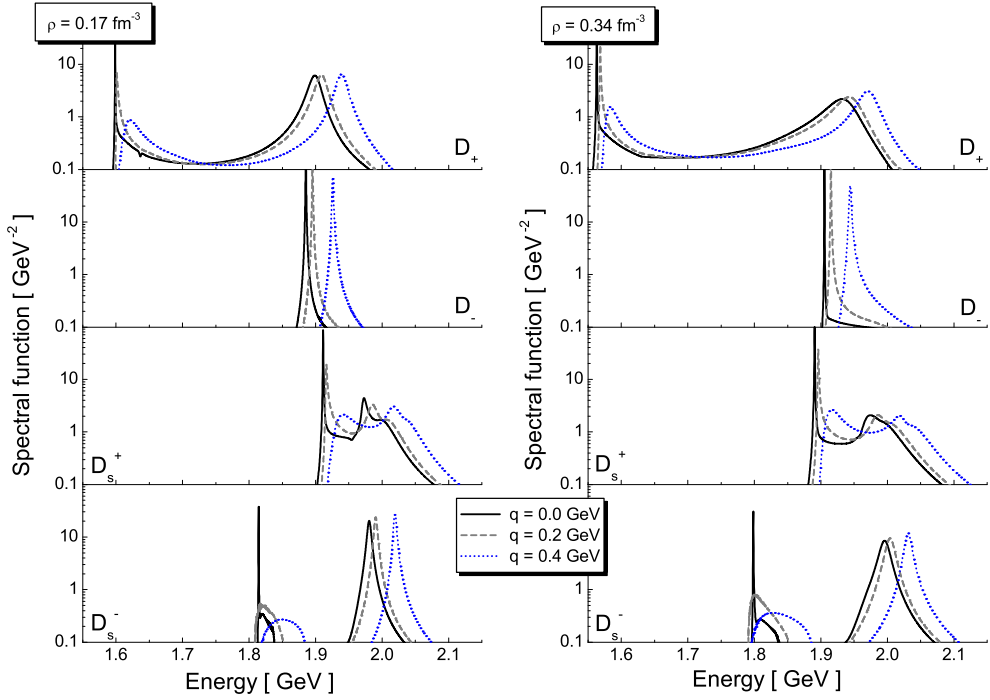}
\end{center}
\caption{Spectral distributions of the $D^\pm$ and $D_s^\pm$ mesons based on the
scattering amplitudes of Fig. \ref{fig:amplitudes}. Results are shown for for meson
momenta 0, 200 and 400 MeV and nuclear densities $0.17$ fm$^{-3}$ and $0.34$ fm$^{-3}$. The self consistent
many-body approach of \cite{Lutz:Korpa:2002} was applied.}
\label{fig:spectral-functions}
\end{figure}

We discuss the free-space scattering amplitudes derived in
\cite{Hofmann:Lutz:05} together with their implications for the in-medium properties of
the D mesons. The results are based on the
self-consistent scheme \cite{Lutz:Korpa:2002} recalled in the previous section.
The s-wave scattering amplitudes are shown in Fig. \ref{fig:amplitudes}. The spectral functions of
the D mesons are shown in Fig. \ref{fig:spectral-functions} for meson momenta 0, 200 and 400 MeV and
nuclear densities $0.17$ fm$^{-3}$ and $0.34$ fm$^{-3}$.

The amplitudes reflect the presence of various resonances. Only the $D^- N$ s-wave scattering process is
not influenced by a resonance. In this case the amplitude is characterized to a large  extent
by the scattering length
\begin{eqnarray}
&& a^{(I=0)}_{D^-N} \simeq-0.16\, {\rm fm} \,, \qquad  \quad a^{(I=1)}_{D^-N} \simeq -0.26\, {\rm fm} \,.
\label{length}
\end{eqnarray}
Given the isospin averaged scattering length the mass shift of a $D^-$ meson in
nuclear matter is fully determined by the low-density theorem \cite{dover,njl-lutz:b}. It holds
\begin{eqnarray}
&& \Delta m_{D^-} = - \frac{\pi}{2}\,\left(\frac{1}{m_N}+\frac{1}{m_D}\right)
\left( a^{(I=0)}_{D^-N}+3\,a^{(I=1)}_{D^-N}\right) \rho + {\mathcal O} \left( \rho^{4/3} \right)
\nonumber\\
&& \qquad \quad \,\simeq  17\,{\rm MeV}\,\frac{\rho}{\rho_0} \,.
\label{}
\end{eqnarray}
The low-density mass shift of 17 MeV is quite close to the self consistent result shown in
Fig. \ref{fig:spectral-functions}. Self-consistency leads to a mass shift of 18 MeV at saturation density
and 38 MeV at twice saturation density.

The positively charged $D^+$ meson interacts with a nucleon in a more complicated manner. The
s-wave scattering amplitude may be characterized roughly in terms of a resonance-pole term and a background
term
\begin{eqnarray}
&& M(\sqrt{s}\,) \simeq  -\frac{|g|^2}{\sqrt{s}-M_R+i\,\Gamma_R/2} + b \,.
\label{Breit-Wigner}
\end{eqnarray}
The isospin-zero amplitude couples strongly to the $\Lambda_c(2594)$ resonance, the
isospin-one amplitude to a so far unobserved $\Sigma_c(2620)$ resonance:
\begin{eqnarray}
&& a^{(I=0)}_{D^+N} \simeq-0.43\, {\rm fm} \,,\qquad \qquad  |g^{(D N)}_{\Lambda_c(2594)}| \simeq 6.6 \,,
\nonumber\\
&& a^{(I=1)}_{D^+N} \simeq -0.41\,{\rm fm}\,,\qquad \qquad |g^{(D N)}_{\Sigma_c(2620)}| \simeq 5.8 \,.
\label{}
\end{eqnarray}

\begin{figure}[b]
\begin{center}
\includegraphics[width=14cm,clip=true]{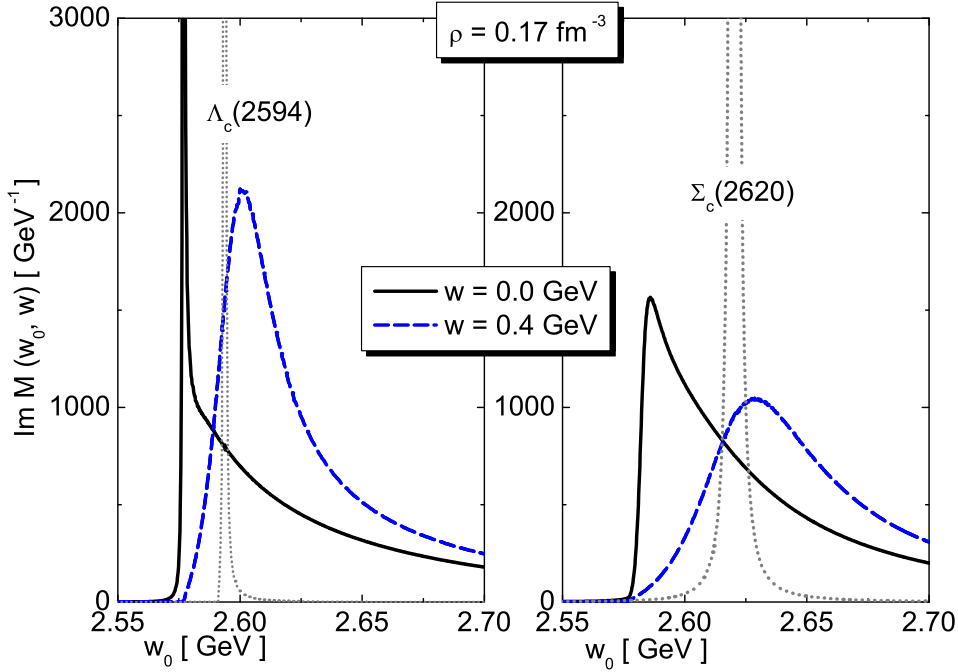}
\end{center}
\caption{Imaginary part of the isospin zero (l.h.p.) and isospin one (r.h.p) $D^+$-nucleon
scattering amplitude at saturation density as compared to the free-space case (dotted lines).
The amplitudes are shown for two values of the resonance three-momentum  $w=0$ GeV and $w=0.4$ GeV.}
\label{fig:resonances}
\end{figure}
In this case the low-density theorem ceases to be useful at densities much smaller than saturation
density. A self-consistent computation is required. Fig. \ref{fig:spectral-functions} demonstrates
that the spectral distribution has a two-mode structure, which is a consequence of important
resonance-hole contributions. Note that the masses of the $\Lambda_c(2594)$ and $\Sigma_c(2620)$ are
almost degenerate. At saturation density the main mode is pushed up by about 32 MeV as compared to
the free-space meson. Nevertheless, due to the resonance-hole state one may expect that
the production of $D^+$ mesons is enhanced in heavy-ion reactions as compared to
nucleon-nucleon collisions. Our results for the $D^+$ meson differ from the previous study
\cite{Tolos:Schaffner:Mishra:2004} significantly. This is a consequence of the quite different
interaction used in the two computations. In particular the work \cite{Tolos:Schaffner:Mishra:2004}
did not predict the isospin one resonance $\Sigma_c(2620)$. The latter dominates the resonance-hole
component in the spectral distribution of the present work. In Fig. \ref{fig:resonances} the spectral
functions of the $\Lambda_c(2594)$ and $\Sigma_c(2620)$ resonances are shown at saturation density as
compared to their free-space distributions.
We observe small attractive shifts in their mass distributions and
significant broadening, at least once the resonances
move relative to the matter bulk.

We turn to the mesons with non-zero strangeness. The study of \cite{Lutz:Kolomeitsev:2004}
predicts the existence of a coupled-channel molecule in the
($D^-_s N,\, \bar D \Lambda,\, \bar  D \Sigma$) system. The latter carries exotic quantum numbers that can not
be arranged by three quarks only. Exotic s-wave states with C=-1 were discussed first
by Gignoux, Silvestre-Brac and Richard \cite{Gignoux:Silvestre-Brac:Richard} and later by
Lipkin  \cite{Lipkin:87}. The binding of about 190 MeV for the $N_{s \bar c}(2780)$ state
predicted in \cite{Lutz:Kolomeitsev:2004} awaits experimental confirmation.
If confirmed the $D^- _s N \to D^-_s  N$
s-wave scattering amplitude must show a prominent pole structure at subthreshold energies. The
amplitude in Fig. \ref{fig:amplitudes} may be characterized with
\begin{eqnarray}
&& a_{D^-_s N} \simeq -0.33\,{\rm fm}\,, \qquad \qquad  |g^{(D^-_s N)}_{N_{s \bar c}(2780)}| \simeq 3.1 \,.
\label{}
\end{eqnarray}
The presence of the $N_{s \bar c}(2780)$ leads to a well separated two-mode structure of the
$D^-_s$ spectral distribution. The main mode is pushed up by less than 10 MeV at nuclear
saturation density.

Recently, it was
pointed out that coupled-channel dynamics predicts attraction also for the $(K\Lambda_c,D^+_s N,  K \Sigma_c)$
system \cite{Lutz:Kolomeitsev:2004,Hofmann:Lutz:05}. An exotic state $N_{\bar c s}(2892)$ about 75 MeV below the
$D_s^+ N$ threshold is predicted. From the  $D^+ _s N \to D^+_s  N$
scattering amplitude of Fig. \ref{fig:amplitudes} we extract the values
\begin{eqnarray}
&& a_{D^+_s N} \simeq-1.21+i\,0.01 {\rm fm} \,, \qquad \qquad |g^{(D^+_s N)}_{N_{c \bar s}(2892)}| \simeq 2.8 \,.
\label{}
\end{eqnarray}
The in-medium spectral distribution of
the $D^+_s$ derived in the self consistent many body approach is most striking. The
two modes expected from the possible existence of the $N_{\bar c s}(2892)$ state
are almost merged into one broad structure, in particular at intermediate meson
momenta 400 MeV. The results are quite analogous to the spectral distribution of the
$K^-$ where the $\Lambda(1405)$ nucleon-hole state gives rise to a broad distribution
\cite{ml-sp,Lutz:Korpa:2002}. Clearly this result is an immediate consequence of the
small binding energy of the $N_{\bar c s}(2892)$ state.

We close with the quest for more detailed studies of the
D-meson nucleon scattering processes to more reliably predict
the in-medium properties of the D mesons.

\end{document}